# An electrospun Polymer Composite with Fullerene-Multiwalled Carbon Nanotube Exohedral Complexes can act as Memory Device

*Fabricio N. Molinari[1], Edwin Barragán[2], Emanuel Bilbao[3,4], Luciano Patrone[3], Gustavo Giménez[3], Anahí V. Medrano[3], Alfredo Tolley[2,5] and Leandro N. Monsalve[3,4]\**

[1]INTI Materiales Avanzados – Av. Gral. Paz 5445 (B1650WAB), San Martín, Bs. As., Argentina.

[2]CNEA-División Física de Metales, Av. E. Bustillo 9500, 8400 S.C. de Bariloche, Argentina.

[3]INTI Micro y Nanotecnologías – Av. Gral. Paz 5445 (B1650WAB), San Martín, Bs. As., Argentina.

[4]CONICET-INTI – Av. Gral. Paz 5445 (B1650WAB), San Martín, Bs. As., Argentina.

[5]CONICET Instituto de Nanociencia y Nanotecnología. Av. E. Bustillo 9500, 8400 S.C. de Bariloche, Argentina.

---

\* Corresponding author Tel: +54 11 4724-6200 ext 6754. E-mail: monsalve@inti.gob.ar

*KEYWORDS:* Conductive composite, Memristor, Polycaprolactone, Carbon nanotubes

*ABSTRACT:* In this work, a novel electrospun conductive polymer nanocomposite made of polycaprolactone with an exohedral complex made of multiwalled carbon nanotubes and fullerene C60 was prepared and characterized. The preparation was straightforward and the complexes self-assembled within the nanocomposite fibers. The nanocomposite showed electrical switching behavior due to charge accumulation of fullerene C60 upon electrical stimulation. Write-once read-many memory devices were fabricated by electrospinning a nanocomposite with 0.8%wt. fullerene C60 onto interdigitated coplanar electrodes. The device retained the ON state for more than 60 days and could be thermally reset, reprogrammed and erased with subsequent electrical and thermal cycling. Moreover, the electrical resistance of the device could be modulated by applying different programming voltage amplitudes and programming times, which revealed its adaptive behavior and potential application to neuromorphic systems.

# 1. INTRODUCTION

Interest in electrically conductive fibers has increased markedly during the last few years in areas such as medicine, sports, military, and energy. These fibers have been applied as power and signal transmitters for ECG measurement[1], strain sensors[2], devices for electrotherapy[3], pressure sensors[4], chemical sensors[5], and photovoltaic devices[6].

Fullerenes are attractive carbon nanomaterials which have promising applications in electronic devices such as thin film memory devices, photovoltaic devices[7–10] and biosensors[11,12]. They possess good electron affinity and behave as electron acceptors, radical scavengers and *n*-type semiconductors. However, they have some drawbacks due to their poor stability and solubility in organic solvents. Several fullerene derivatives have been synthesized in order to improve stability and processability[13–16].

Hybrid nanostructures made of carbon nanotubes (MWCNT) and fullerenes have been prepared and studied. It has been demonstrated that both carbon nanomaterials can associate and form stable nanostructures of MWCNT-C60 endohedral and exohedral complexes[7–10,15,17–19]. These structures combine the ability of C60 to capture electrons and the charge transport properties of MWCNT. Their properties have impact in avoiding recombination of electron-hole pairs in bulk heterojunction photovoltaic devices, leading to improvement of efficiency in these devices[7–10]. Charge transport properties of these structures are affected by chemical functionalization. According to Li *et al.*[8], short electron-withdrawing groups on the surface of MWCNT serve as electrical connection between them and C60 whereas long alkyl chains act as an insulating barrier that blocks their charge transport ability.

Polycaprolactone (PCL) is a semicrystalline, thermoplastic, biocompatible and biodegradable polymer. It is easily processable due to its solubility in various solvents and its low melting point (60 °C). Moreover, its mechanical properties make it attractive for the preparation of flexible electrospun fibers and polymer-based composite materials.

The electrospinning technique allows the fabrication of sub-micron structures (either fibers or particles) from melt or solution-processed materials by means of an electric field applied between a source (usually a needle) from which the material flows, and a collector. It was first published in 1934[20], but it was not until the last decade that the technique became popular for materials research and production[21]. This technique can be used for the fabrication of fibers made of polymers, polymer composites and inorganic materials. Moreover, morphology and alignment of the fibers can be controlled for tuning the desired properties for the mat (lightweight, porosity, selective location and alignment of different materials within the fiber, etc.). This technique has already been tested for controlled release applications[22], sensor fabrication[12,23] and electronic devices such as field effect transistors[24] and photoactive materials for photovoltaics[25].

Specially, electrospinning has been employed for the preparation of nanofibrous PCL/MWCNT functional electrospun composite materials with conductive[26] and controlled release properties[22]. Moreover, composite materials made of PCL with covalently-linked fullerene C60 have been prepared and processed by electrospinning[15,27]. The distinctive properties of these composites that have been described so far are the nucleating effect of C60 in PCL and their ability to generate singlet oxygen. The use of C60 along with MWCNTs in electrospun polymer composited has been scarcely explored[25]. Moreover, there are not many reported applications of electrospun materials for non-volatile memory devices[28].

The basic goal of a memory device is to provide a means for storing and accessing binary digital data sequences of "1's" and "0's", as one of the core functions (primary storage) of modern computers. According to the storage type of the device, electronic memory can be divided into two primary categories: volatile and non-volatile memory. A non-volatile memory can store data even if it is disconnected from the power supply. There are two types of non-volatile memories: Non-volatile random access memory (NVRAM) and write-once read-many memory device (WORM). NVRAM can be written an erased many times whereas WORM is a non-volatile memory device that can be used to store archival standards, databases and other massive data where information has to be reliably preserved for a long period of time[29].

In conventional silicon-based electronic memory, data are stored based on the amount of charge stored in the memory cells. Organic/polymer electronic memory stores data in an entirely different way, for instance, based on different electrical conductivity states (ON and OFF states) in response to an applied electric field. Different mechanisms for organic memories have been illustrated such as charge transfer, conductive filament formation and charge-trapping[30]. Organic/polymer electronic memory is likely to be an alternative or at least a supplementary technology to conventional semiconductor electronic memory, and possesses many advantages such as large-area processability, flexibility and low cost.

Regarding memory devices, fullerene C60 and some of its derivatives such as [6,6]-phenyl-C61-butyric acid methyl ester (PCBM) have been employed alone and in the form of thin-film polymer composites for the fabrication of non-volatile memories. This nanomaterial has been used in different memory architectures such as metal oxide-semiconductor capacitors[31,32], charge trapping memory transistors[33,34] and memristors[35–46]. Memristors are particularly attractive because they mimic brain synapses. Biocompatible memristor devices are highly desirable due to their potential application in biomedical electronics.

Adaptive devices that mimic neural synapses are of interest due to their potential application to low-power neuromorphic systems and integration to biological neural networks. These devices show stimuli-dependent response (potentiation or depression) that allow strengthening or weakening of different connections in a given circuit. Such devices have been recently fabricated using a C60-polymer composite[33,34]. Their adaptive behavior has been attributed to the dynamic trapping/ de-trapping process of electrons and holes within the composite material.

In this work we report the preparation and characterization of a conductive electrospun composite made with a biodegradable polyester and C60/MWCNTs complexes with electrical switching behavior. The unique characteristics of the composite were applied for the fabrication of a non-volatile memory device with coplanar electrodes that could be reset by thermal annealing.

## 2. EXPERIMENTAL SECTION

**2.1. Materials.** PCL (Mw 80000, Sigma-Aldrich), Fullerenes C60 (99.9%, Sigma-Aldrich), polyvinyl pyrrolidone PVP K30 (Anedra, Argentina), MWCNT(Nanocyl 7000, Belgium), xylene, toluene, hexane, DMF, and acetone were reagent grade and used straight from the bottle. TLC plates (Silicagel 60 F254 on aluminum) were from Merck (Germany).

**2.2. Preparation of Polymer Solutions.** Fullerene C60 (35 mg) was dissolved in xylene (11 ml) in ultrasound bath at 40 °C for 90 minutes. Then 2.7 g of PCL was added to 10ml of the solution and dissolved at 50 °C under magnetic stirring. The resulting solution was mixed with 10 g of a dispersion of MWCNTs in DMF (0.7%wt.) prepared according to a previously described procedure[26]. Once homogeneous additional gram of PCL was dissolved under magnetic stirring for 3 h and used immediately.

**2.3 Electrospinning conditions and Device fabrication.** An Y-flow electrospinner 2.2.D-500 (Y-flow SD, Spain) was used for electrospinning. Distance to the collector was optimized in order to obtain regular and dry fibers. Different tests led to determine 26 cm as the optimum distance between the needle and the collector. The flow rate of the solution was 1 ml.h$^{-1}$ for all samples. A rotary drum collector was employed in order to produce mats of aligned fibers. The setup included two high voltage sources: one at the needle between of +6 and +12 kV and the other at the collector between of -15 and -17 kV. Preliminary tests showed that fiber alignment increased with rotation speed. However, rotation speeds above 1000 rpm induced turbulent air stream that threw the fibers out of the collector.

PCL-MWCNT/C60 solutions were spun onto interdigitated gold electrodes of different width to length (W/L) ratio (L=1-100 μm, W/L=50-2000) patterned on a Si/SiO$_2$ (300 nm) substrate for 15 minutes using a rotary collector in order to align the fibers perpendicular to electrode fingers. The dies with the deposited fibers were annealed at 60 °C for 20 minutes in order to improve contact between fibers and electrodes.

**2.4. Characterization.** Scanning electron microscopy and Focused Ion Beam (FIB) experiments were performed in a Helios Nanolab 650 (FEI). FIB was employed for lamellae preparation, mounting and polishing from composite fibers (see Supporting information for details). The obtained piece (final thickness: ~30 nm) was welded to a FIB-TEM copper grid and analyzed by High resolution transmission electron microscopy in a Tecnai F20 UT microscope, operated at 200 kV. Fiber diameter and alignment measurements were carried out by analysis of SEM images using Image J software[47]. DSC was performed in a TA Q2000 from -90 to 100 °C and from 100 to -90 °C at 10 °C/min. Raman microscopy was performed using a Renishaw In via Raman Microscope with 785 nm laser source. Electrical characterization was performed using a Keithley 4200 SCS equipped with a manual probe station (Wentworth Lab AVT 702). X-Ray Diffraction was performed using a diffractometer Philips pw 1730/10.

**2.5. Extraction of C60/MWCNT complexes.** C60/MWCNT complexes were extracted in acetone. A small sample of PCL-C60/MWCNT fibers was dissolved in acetone with magnetic stirring. A black-colored dispersion was obtained and then centrifuged in order to separate the precipitate. The precipitate was washed and centrifuged 3 times with acetone in order to wash the complexes and eliminate the PCL.

## 3. RESULTS AND DISCUSSION

**3.1. Fiber characterization.** SEM images of devices were taken in order to characterize fiber morphology (Figures 1a and 1b). These images revealed a monolayer of fibers aligned perpendicular to electrode fingers (average angle: 89° respect to electrode fingers). Fiber diameter was approximately 650 nm and their surface was regular and smooth. Some MWCNT entanglements could also be noticed on the fibers. It was also observed that the thermal treatment performed in order to optimize electrical properties of the device did not affect the fibrous structure of the electrospun layer.

Lamellae prepared by FIB thinning of the PCL-C60-MWCNT fibers were analyzed with HRTEM. In such specimens the MWCNTs were visible in underfocussed or overfocussed images, but the strong contrast from the amorphous PCL hindered the observation of the C60 fullerenes. An example of such images is shown in Figure 1c. Instead, in images of extracted C60/MWCNT complexes the association of C60 with MWCNTs was observed. Figure 1d shows a HRTEM image of an extracted C60/MWCNT complex, where C60 fullerenes can be observed attached to the outer surface of the MWCNT. The condition is overfocussed to enhance the contrast of the C60 fullerenes. The C60 fullerenes are best observed on the external edges of the MWCNT, although some can also be observed superposed to the contrast of the walls of the MWCNT. The diameter of the MWCNTs ranged from about 10 to 19 nm, and the C60 fullerenes were typically of about 1 nm in diameter.

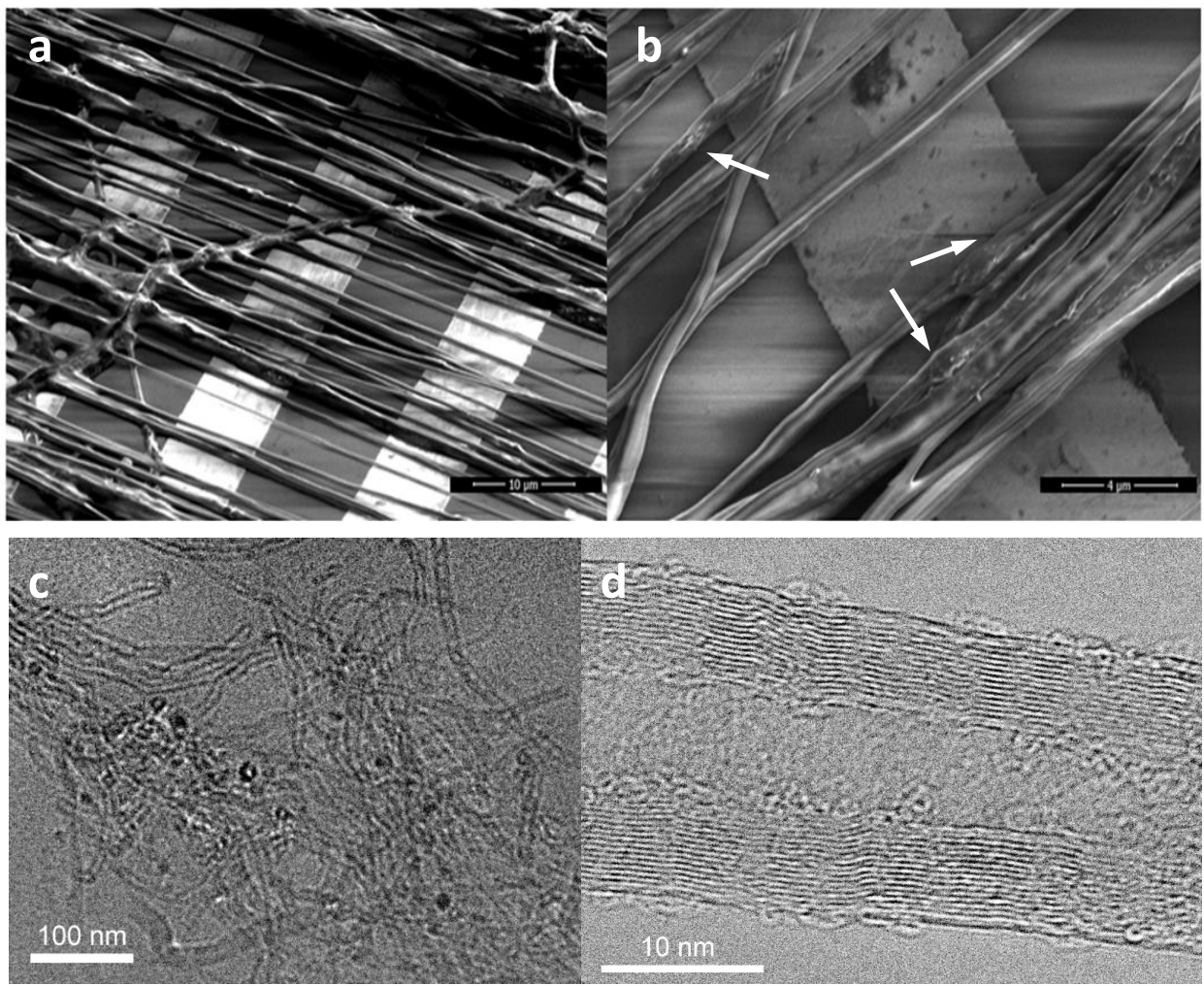

**Figure 1.** Electron microscopy images of PCL-C60/MWCNT fibers showing a) parallel electrospun fibers onto interdigitated electrodes after the thermal treatment, b) MWCNT entanglements within PCL-C60/MWCNT fibers (pointed by arrows), c) a MWCNT entanglement within a single FIB-thinned PCL-C60/MWCNT fiber and d) C60 fullerenes attached to the external surface of a single MWCNT.

SEM images showed a combination of well-dispersed MWCNTs along with and aggregates that ensures electrical percolation of the electrospun fibers according to our previous studies[26]. Moreover, HTREM experiments on the extracted C60/MWCNT complexes showed that C60 fullerenes were attached to the outer surface of the MWCNTs.

Raman spectra of the fibers were acquired in order to determine intermolecular association between PCL, MWCNT and C60. Spectra of fibers prepared with PCL alone, PCL with MWCNTs PCL with C60 and PCL with C60/MWCNT under the same conditions were compared to each other (Figure 2).

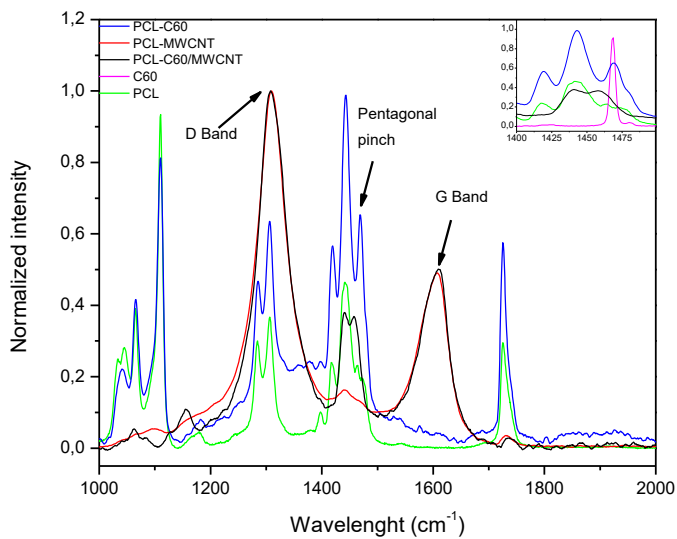

**Figure 2.** Raman spectra of PCL-C60 (blue), PCL-MWCNT (red), PCL-C60/MWCNT (black) and PCL fibers (green). D band and G band of MWCNTs and pentagonal pinch mode of C60 are indicated. The inset shows the pentagonal pinch in fibers containing C60 compared to pristine C60 (magenta).

Three bands corresponding to $\delta(CH_2)$ vibration of PCL were present between 1410 and 1490 cm$^{-1}$ in the spectrum of PCL fibers. Additionally, the spectrum of fibers with C60 showed an overlapped band at 1470 cm$^{-1}$ which corresponds to the pentagonal pinch of C60. Interestingly, an overlapped band could be also observed in the spectrum of the fibers with C60 and MWCNTs, but downshifted about 10 cm$^{-1}$.

Regarding the bands corresponding to MWCNTs no significant change in the relative intensity or position of the D and G bands was observed upon the addition of C60.

These results suggest the occurrence of strong interactions between C60 and MWCNTs in the electrospun fibers. Therefore, an extraction procedure of C60/MWCNT complexes from the fibers was performed. Extraction helped to eliminate the interference of PCL in Raman spectra. Raman spectra were recorded from the extracts (Figure S4). In this case, PCL could be completely eliminated and up to six signals from C60 could be identified along with D, G and D* bands of MWCNTs. The bands from C60 are described in Table 1. A clear shift to lower wavenumbers was observed for these bands in the extract and the downshift of the pentagonal pinch was confirmed.

**Table 1.** Raman shifts of pristine C60 and MWCNT/C60 complexes extracted from PCL-C60/MWCNT electrospun fibers.

| C60 – Freq. (cm$^{-1}$) | Extract – Freq. (cm$^{-1}$) | Downshift – Freq. (cm$^{-1}$) |
|:---:|:---:|:---:|
| 1469 | 1464 | 5 |
| 772 | 770 | 2 |
| 709 | 703 | 6 |
| 495 | 491 | 4 |
| 432 | 427 | 5 |
| 271 | 267 | 4 |

Regarding the downshift of the pentagonal pinch mode of C60, it has been reported previously as an indication of π-backbonding of C60 with metal complexes[48] and doping of C60 with negative charges [49]. In C60/MWCNT complexes, a downshift in the pentagonal pinch mode has been reported as indicative of a cross coupling reaction between derivatives of both nanomaterials[50] and encapsulation of C60 into single-walled carbon nanotubes (SWCNTs)[51]. On the other hand, Yekymov et al.[52] reported no changes in the Raman spectra of carbon nanotubes and C60 after the formation of exohedral complexes. In order to define if C60 could be effectively separated from the exohedral complex a thin layer chromatography was run on the extract (Figure S5). We found that C60 could be separated in its original form from the complex. Therefore, the changes in the Raman shifts could be only explained by the formation of a C60/MWCNT exohedral complex.

Regarding lack of changes in the MWCNT Raman spectra, Yang et al.[53] described an upshift of the G band after encapsulating C60 into SWCNTs as a sign of charge transfer between both carbon nanomaterials. In our case, this effect could be minimized considering the fact that we used MWCNTs instead of SWCNTs and thus a much smaller fraction of the graphene layers are exposed to C60.

DSC and XRD experiments were performed in order to elucidate differential effects of MWCNTs and C60 in the PCL crystalline domains.

Nucleation efficiency (*NE*) is one of the best ways to quantitatively assess the nucleation capacity of an additive to act as nuclei for the crystallization of a given polymer compared to self-nucleation. The self-nucleation scale for PCL-MWCNT composites was employed as used by Pérez et al.[54] The nucleation efficiency can be calculated according to the following expression:

$$NE = \frac{T_{c,NA} - T_{c,PCL}}{T_{c,max} - T_{c,PCL}} \times 100 \qquad (1)$$

where $T_{c,NA}$ is the peak $T_c$ value determined in a DSC cooling run for a sample of the polymer with the nucleating agent (NA), $T_{c,PCL}$ is the peak $T_c$ value for neat PCL, and $T_{c,max}$ is the maximum peak crystallization temperature determined after neat PCL has been self-nucleated at the ideal self-nucleation temperature or $T_{s,ideal}$ (i.e., the self-nucleation temperature that produces maximum self-nucleation). The values of $T_{s,ideal}$ and $T_{c,max}$ were determined according to Trujillo et al.[55] (See Supporting information for experimental details and DSC plots). Calculated crystallinity ($\chi$), $T_c$ and NE of of fibers prepared with PCL alone, PCL-MWCNT, PCL-C60 and PCL-C60/MWCNTs from DSC experiments are shown in Table 2.

NE was found to be higher than 100% for every composite fiber, which indicates supernucleation. However the presence of C60 reduced NE compared to PCL-MWCNT, even in combination with MWCNT.

**Table 2.** $T_c$, $\chi$ and NE for PCL, PCL-C60, PCL-C60/MWCNT and PCL-MWCNT fibers.

| Sample | $T_c$ (°C) | $\chi$ (%) | NE (%) |
|---|---|---|---|
| PCL | 26.5 | 65.6 | - |
| PCL-C60 | 29.9 | 43.3 | 112 |
| PCL-C60/MWCNT | 33.5 | 41.3 | 146 |
| PCL-MWCNT | 42.3 | 46.3 | 228[26] |

It is well known that both carbon nanotubes[26,54,55] and fullerenes[14] can act as nucleation agents in polymer composites. DSC analysis revealed that MWCNTs and C60 served as nucleation spots for PCL, increasing its NE. However, the effect of raising NE of both nanomaterials was not additive: *NE* for PCL-C60/MWCNT fibers remained similar to PCL-C60 fibers and much lower than for PCL-MWCNT fibers. This indicates that the nucleating effects of MWCNTs were hindered by C60 due to unavailability of the surface of the former for the nucleation of PCL.

XRD diffractograms of electrospun fibers showed the characteristic peaks for (111) and (100) lattice planes of PCL (Figure 3). XRD analysis revealed a downshift of the peak observed at a 2θ=21.4°, which corresponds to (111) lattice plane of PCL crystal, when C60 (-0.26°) and MWCNTs and C60 (-0.18°) were incorporated. This effect was not observed when only MWCNTs were incorporated to the fibers. Using XRD it was possible to calculate the effect of the presence of carbon nanotubes and fullerenes in the crystal thickness of PCL using the following equation[56]:

$$t = \frac{K\lambda}{B\cos\theta_B} \qquad (2)$$

Where K is a dimensionless constant that may range from 0.89 to 1.39, depending on the specific geometry of the scattering objects. For a perfect two-dimensional lattice, where every point on the lattice emits a spherical wave, numerical calculations yield a lower bound of 0.89 for K. A cubic three-dimensional crystal is best described by K = 0.94, whereas analytical calculations for a perfectly spherical object yield K = 1.33. K was taken as 0.9 in this study. B is a measure of the peak width, the full width at half-maximum at 2θ. Table 3 shows the calculated crystal thickness of PCL for pristine PCL, PCL-MWCNT, PCL-C60 and PCL-C60/MWCNT electrospun fibers.

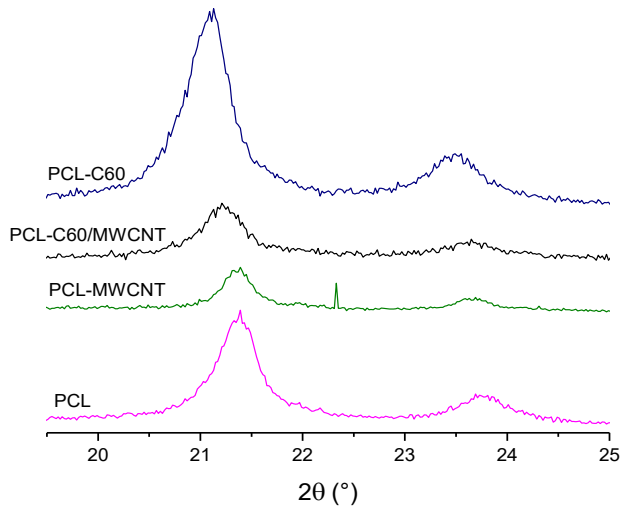

**Figure 3.** XRD difractograms for PCL, PCL-C60, PCL-MWCNT and PCL-C60/MWCNT fibers. A downshift for the characteristic peaks of (111) and (100) lattice planes of PCL was observed upon the addition of MWCNT, C60/MWCNT and C60.

**Table 3.** Characteristic peaks of (111) lattice plane of PCL and calculated crystal thickness of PCL for PCL, PCL-MWCNT, PCL-C60 and PCL-C60/MWCNT.

| Sample | 2θ(°) | Crystal thickness(nm) |
|---|---|---|
| PCL | 21.4 | 22.4 |
| PCL-MWCNT | 21.4 | 27.1 |
| PCL-C60/MWCNT | 21.2 | 21.2 |
| PCL-C60 | 21.1 | 19.0 |

The estimation of the crystal thickness of PCL served also for explaining the presence of the exohedral complex. The presence of MWCNTs stimulated the growth of larger PCL crystals while the presence of C60 produced smaller crystals. This behavior could be explained by geometrical issues. The growth of larger PCL crystals is more difficult in presence of C60 because the greater curvature of fullerenes and their spherical geometry. The PCL in fibers containing C60/MWCNT complexes showed a crystal thickness close to those in fibers containing only C60, which suggests that most of the PCL crystalizes on the surface of C60. These results are consistent with DSC results and with the presence of the proposed hierarchical structure of carbon nanotubes and fullerenes.

Intermolecular interactions between PCL, C60 and MWCNTs were revealed by Raman, DSC and XRD experiments. Along with HRTEM analysis we are able to propose a hierarchical structure made of MWCNT, C60 and PCL within the electrospun fibers (Figure 4). MWCNTs nucleate the C60 molecules forming a core-shell structure.

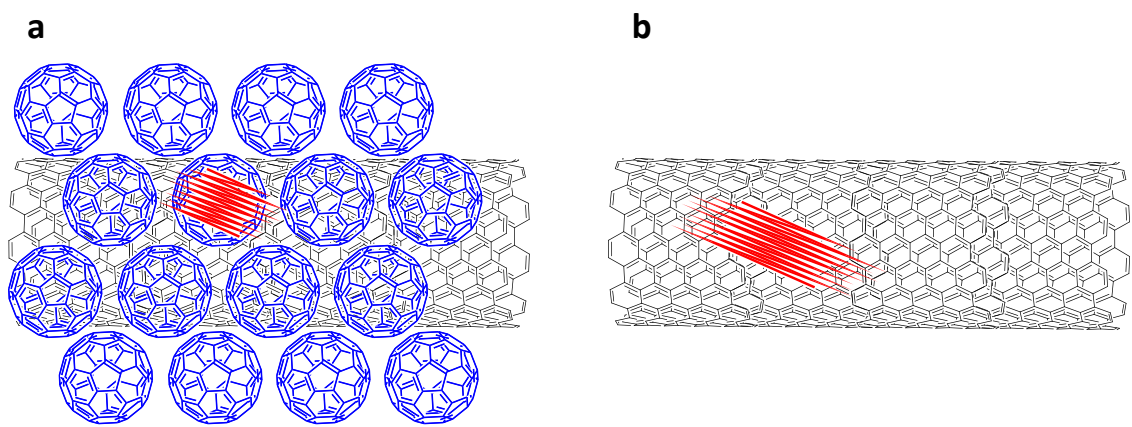

**Figure 4.** Qualitative diagram of a hierarchical structure composed of a core of MWCNTs (black) and a shell of C60 (blue) surrounded by PCL crystals (red) on the surface of the C60 shell (a). When C60 is not present, PCL is able to crystallize as thicker crystals on the MWCNT surface (b).

**3.2. Electrical properties of fiber-based memory devices.** Electrical properties of nanocomposite fibers were performed by applying different stimulation voltages between interdigitated gold microelectrodes with the electrospun fibers deposited on top. Voltage sweeps between -0.2 and 0.2 V were performed in order to determine the electrical resistance of the devices before and after stimulation as the inverse of the slope of the IV plots. Within this range the fibers showed linear ohmic behavior. Surface resistivity of the fibers could be estimated as approximately 4.94 x $10^8$ Ω/sq. The addition of C60 with 1.8%wt. MWCNT contributed to a dramatic increase of fiber conductivity compared to the addition of 1.8%wt. MWCNT only (surface resistivity: 2.76 x $10^{13}$ Ω/sq[26]).

MWCNT concentration was adjusted to ensure conductivity near the percolation threshold. Fewer amounts of MWCNTs impaired electrical conduction whereas higher MWCNT loads minimized the resistive switching effect of C60. Moreover, fiber alignment perpendicular to electrode fingers minimized the resistivity of the devices. The presence of C60 also enhanced the conductivity of the fibers compared to MWCNT alone.

Further voltage sweeps at higher potentials leaded to a decrease in the resistivity of the fibers. Aiming to define the switching threshold voltage, voltage sweeps starting from 0 V with a step of 1 V between sweeps were performed and the electrical resistance was measured after each cycle (Figures 5a and b). The devices showed an incremental behavior in conductivity from 5 V up to 12 V. Once switched on, the devices showed high stability retaining their electrical resistance for at least 600s at programming voltage (Figure 5c). Larger stimulation voltages leaded to the breakdown of the device. Moreover, successive programming and reading of devices at constant voltage was applied in time at different programming voltages (Figure 5d). In this case, device switching at 3 and 4 V could be achieved in two steps after up to 60 s of stimulation cycles. Higher programming voltages leaded to switching in a single step after a single programming cycle of 60 ms. However, the final $R/R_0$ value of the device turned to be both programming time and voltage-dependent, which revealed the adaptive characteristics of the device.

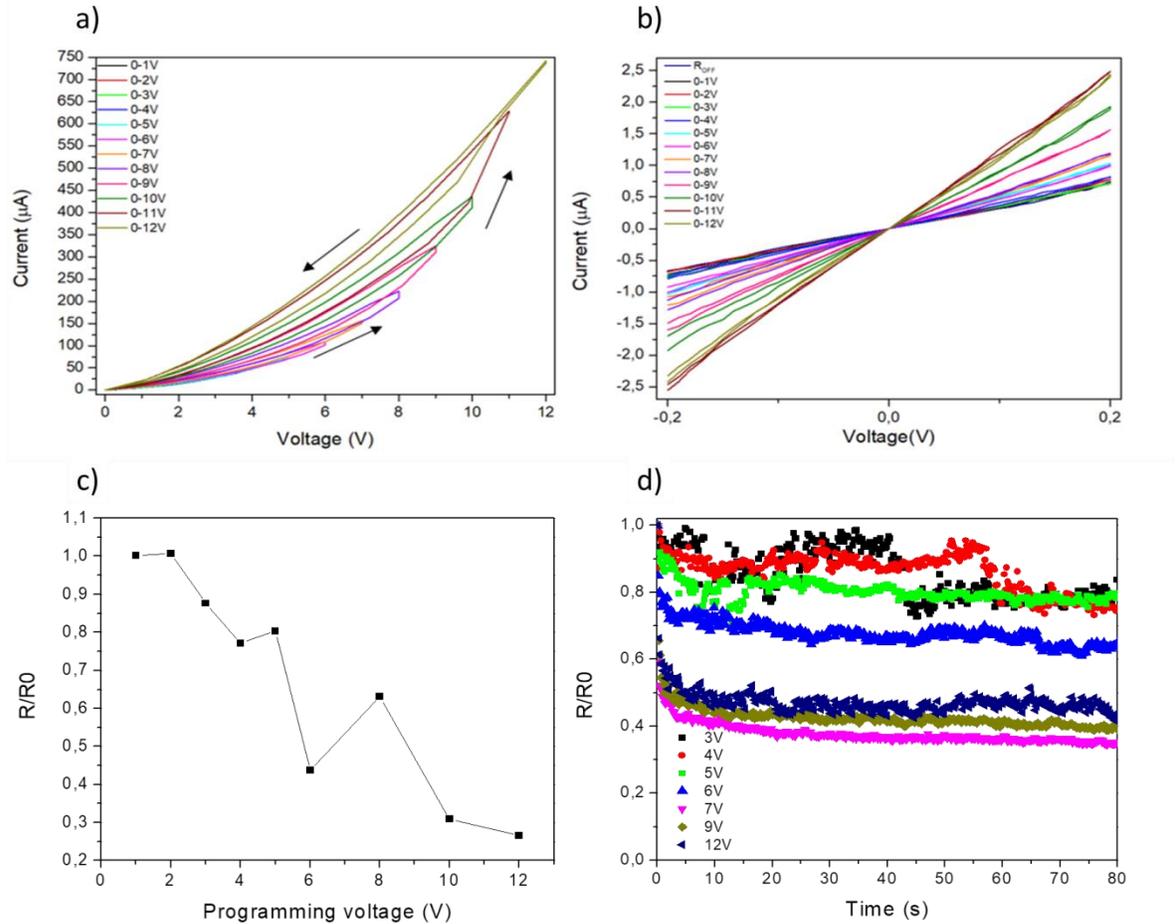

**Figure 5.** (a) Electrical stimulation of memory devices starting from 0 V with a step of 1 V between voltage sweeps and (b) IV curve after each stimulus in the linear region. (c) $R/R_0$ at 200 mV vs. programming voltage after 600 s at programming voltage. (d) $R/R_0$ at 200 mV vs. time after successive programming pulses (60 ms) at different programming voltages.

Figures 6a and b show the typical switching behavior of the device. Stimulation with voltage sweeps to 12 and -12V revealed a clear bistable behavior. Reverse sweeps failed to reset the device and the electrical resistance of the ON state remained almost constant during 10 forward/reverse voltage cycles. Such behavior is typical for WORM-type devices. The calculated $R/R_0$ for the devices varied from 0.25 to 0.7 depending on the device. In order to assess the stability of the device the electrical resistance was measured during the following days after the device has been switched on (Figure 6c). $R/R_0$ ratio increased during the first 20 days after the device was switched on. After this point no changes were observed for at least 60 days.

In order to determine the influence of the polymer matrix in the relaxation of the ON state, an annealing of the devices at 60°C was performed after switching them on. It was found that thermal annealing reset the devices back to the OFF state. Figure 5d shows that the devices could be cycled at least 6 times by successive thermal and electrical stimuli.

In order to evaluate the influence of the electrospinning process in the final characteristics of the device, devices were fabricated by spin coating using the same PCL-C60/MWCNT dispersion used for electrospinning onto interdigitated electrodes (Figure S8). Contrary to electrospun devices, spin-coated devices showed an increase of electrical resistance (Figure S9). Spin-coated films were also characterized by Raman spectroscopy which showed no downshift of the pentagonal pinch mode (Figure S10). Therefore, the electrospinning process played a fundamental role in the formation of the C60/MWCNT exohedral complexes.

Resistive switching behavior of polymer composites containing C60 has been previously described. Composition characteristics and type of device are resumed in Table 4. As in previous works describing memory devices fabricated with polymer composites containing C60 the resistive switching behavior can be explained by charge accumulation of C60. Therefore, the increment in conductivity of PCL-C60/MWCNT fibers with increasing the amplitude of the programming pulse below 12 V indicates enhanced charge trapping ability from C60/MWCNT complexes with increasing voltage. A similar behavior has been described for transistor memory devices having a C60 polymer composite as charge trapping layer[33,34]. Moreover, the adaptive behavior of the devices could be explained by the conduction between C60/MWCNT complexes within the composite fibers by hopping and mediated by the charge trapping ability of C60 at the interface of MWCNTs. This is an important feature regarding the potential application of these fibers made with a biocompatible polymer in artificial synapses.

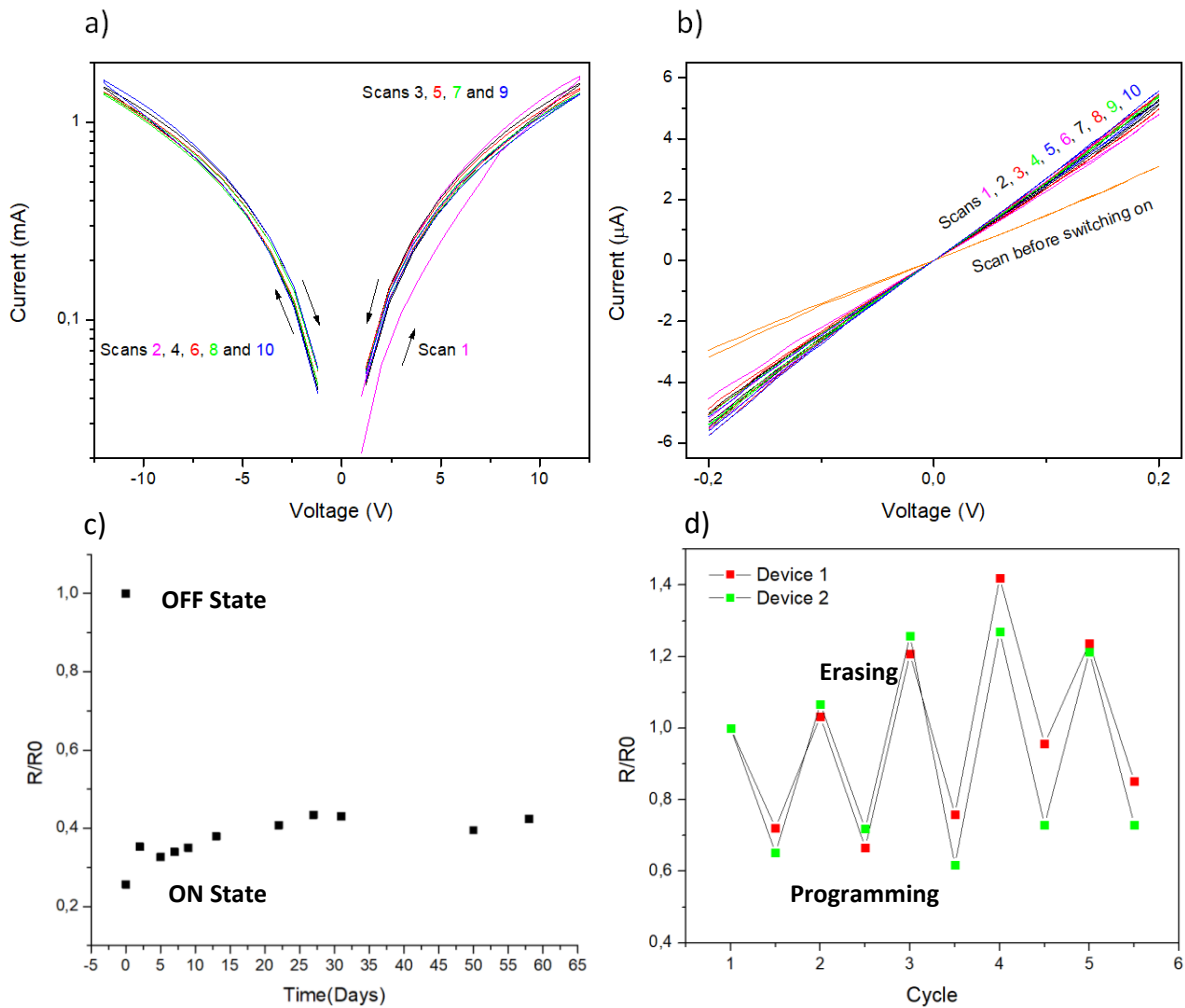

**Figure 6.** a) Successive programming cycles of memory device, b) IV curves showing differences in electrical resistance after successive programming cycles, c) evolution of electrical resistance of memory device in time after a programming cycle and d) electrical resistance variation of memory devices under electrical programming and thermal erasing cycles.

In our case, the use of C60/MWCNT exohedral complexes instead of C60 alone allowed a dramatic reduction of C60 content required to achieve a WORM-type device behavior. In our device, the conduction path between vicinal MWCNTs was enhanced due to charge accumulation of C60 in their interface. Reported WORM-type polymer composite devices need at least 7.5%wt whereas 0.8%wt C60 is enough for displaying such behavior in our electrospun PCL-C60/MWCNT fibers. Moreover, these reported devices display sandwich-type electrode configuration whereas our device has coplanar electrodes, which simplified the fabrication process.

**Table 4.** Reported characteristics of Polymer-Fullerene memristor devices.

| Type of device | Matrix | Fullerene type | Fullerene content (%wt.) | Reference |
|---|---|---|---|---|
| WORM | Syndiotactic PMMA | C60 | 20% | [39] |
| WORM | PCL | C60 | 0.8% | This work |
| NVRAM | PS | C60 | 5% | [38] |
| NVRAM | PS | C60 | 5% | [37] |
| WORM | PS | C60 | 7.5-20% | [37] |
| NVRAM | Natural rubber | C60 | 0.1-10% | [43] |
| NVRAM | Nylon | C60 | 5% | [57] |
| NVRAM | Fullerene-based polymer | Covalently-linked C60 | >37% | [36] |
| NVRAM | Polyvinylphenol | C60 | 5% | [35] |
| NVRAM | P3HT | C60 or PCBM | 5% | [41] |
| NVRAM | PS-b-PMMA | PCBM | 0.05% | [45] |
| WORM | P3HT | PCBM | 50% | [40] |

Regarding stability of the ON state in time, it could be hypothesized that charge accumulation induces electrostatic repulsion between neighboring C60/MWCNT complexes. This force can lead to an increase of the interparticle distance within the polymer matrix, which explains an increase of resistance during the days after switching on the device. It is possible that particle movement provokes plastic deformation in PCL at nanoscopic level, which could explain the observed relaxation time. Another probable explanation for resistive relaxation is the occurrence of accumulated strain within the fiber composite during the electrospinning process.

Thermal annealing of the devices revealed the role of the polymer matrix in relaxation, possibly by accelerating charge recombination between C60 and MWCNT when heated up to the melting point of PCL. Thermal reset of WORM-type devices has been described previously for the case of organic donor-acceptor layers[30]. Device cycling could be achieved due to the formation and breakage of new conduction paths in each cycle. It could be also noticed that resistance of the ON and OFF states after electrical programming and thermal erasing cycles showed variations with a trend in increasing resistance of the OFF state. This can be attained to an adaptive behavior of the composite material by the dynamic formation and breakdown of conductive paths depending of its thermal and electrical history.

## 4. CONCLUSIONS

A novel nanocomposite material made of PCL and exohedral C60/MWCNT complexes was processed by electrospinning for the first time. The noncovalent interaction of C60 on the surface of MWCNTs was confirmed by Raman spectra, DSC and DRX experiments and HRTEM. The material showed increased conductivity compared to PCL-MWCNT electrospun fibers and resistive switching behavior. This feature was used for the fabrication of a WORM-type memory microdevice by electrospinning PCL-C60/MWCNT fibers onto interdigitated gold microelectrodes. Compared to previously reported fullerene-based WORM-type devices, lower content of C60 (0.8%wt.) was sufficient to achieve this resistive switching behavior. Once switched on, the device retained a low resistance state for more than 60 days, albeit increasing its resistance due to relaxation of accumulated strain within the fibers due to charge accumulation. The device could be reset by thermal annealing at 60°C. Further electrical and thermal cycling of the device proved that it could retain its resistive switching properties.

In conclusion, this paper reported the preparation and characterization of a novel electrospun nanocomposite made of a carbon nanocomplex and a biocompatible and biodegradable polymer. Moreover, the nanocomposite proved to be responsive to electrical and thermal stimuli, which allowed the fabrication of an adaptive resistive switching microdevice and paves the road to the development of flexible responsive devices that could interact with biological media.

**ACKNOWLEDGEMENTS**

We thank INTI, ANPCyT (PICT 2013-0427, PICT 2014-3748 and PICT 2017-2787) and CONICET (PIP 11220150100967) for financial support. Authors are also grateful to Sandra Amore, Natalia Loiacono and Rodrigo Alvarez for the XRD experiments, Eliana Mangano for the fabrication of electrodes and Sandra Jung for the statistical analysis on fiber diameter and angle. LNM and AT are research staff of CONICET.

**SUPPORTING INFORMATION**

**1. Lamellae preparation for TEM experiments.** In order to obtain HRTEM images of the nanocomposite electrospun fibers, thin film lamellae were prepared by FIB microscopy. A protective layer of 10x2 μm of Pt (thickness ~150 nm) was deposited with the ion beam accelerated at 30 KV upon an individual fiber. After that, two trenches of 10 μm in depth were made at each side of the protective layer. The lift-off of the lamella was carried on by welding the Pt layer to a tungsten tip attached to a micromanipulator. Finally, the obtained lamella was welded to a FIB-TEM copper grid in order to perform a thinning of the polymer nanocomposite layer to a final thickness of about 30 nm, with pin-holed zones which served for the observation of the carbon nanofillers by HRTEM.

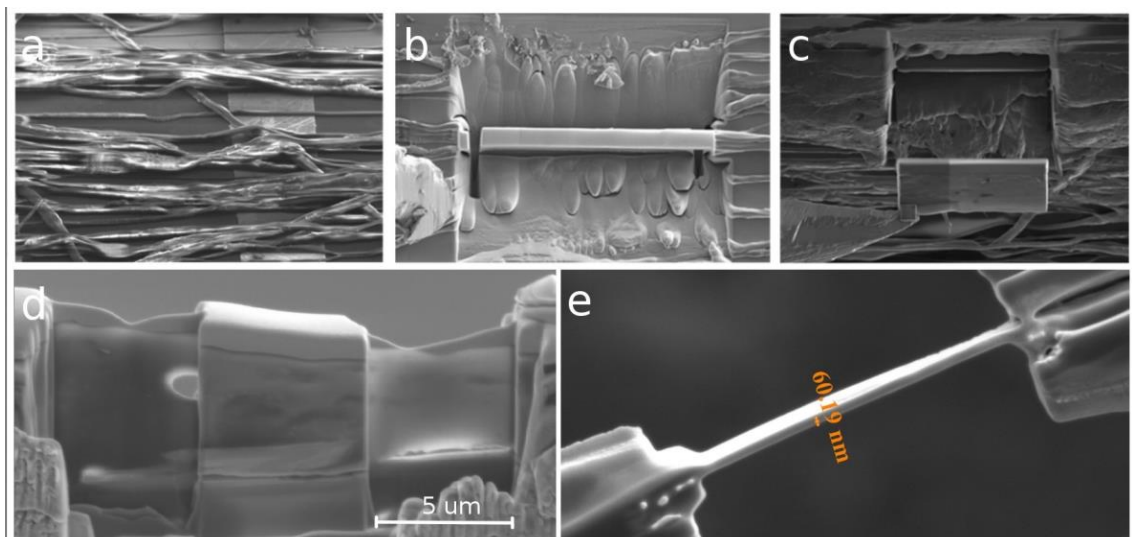

**Figure S1.** Lamella TEM Preparation process: a) fiber monolayer sample with the target area to make a HRTEM characterization, b) Pt deposition and trench milling, c) lamella lift-off with W tip, d) sample attached by Pt welded to a FIB/TEM copper grid and e) thinning to final thick of about 60 nm.

**2. Statistics on fiber diameter and orientation.** A SEM image (Figure S2) was analyzed with Image J software to determine the average fiber diameter and fiber orientation. Diameter and angle respect to the electrode fingers distribution were analyzed with Minitab and the results are shown in Figure S3. Mean diameter and mean angle are shown in Table S1. Both analyses were performed through 120 measurements.

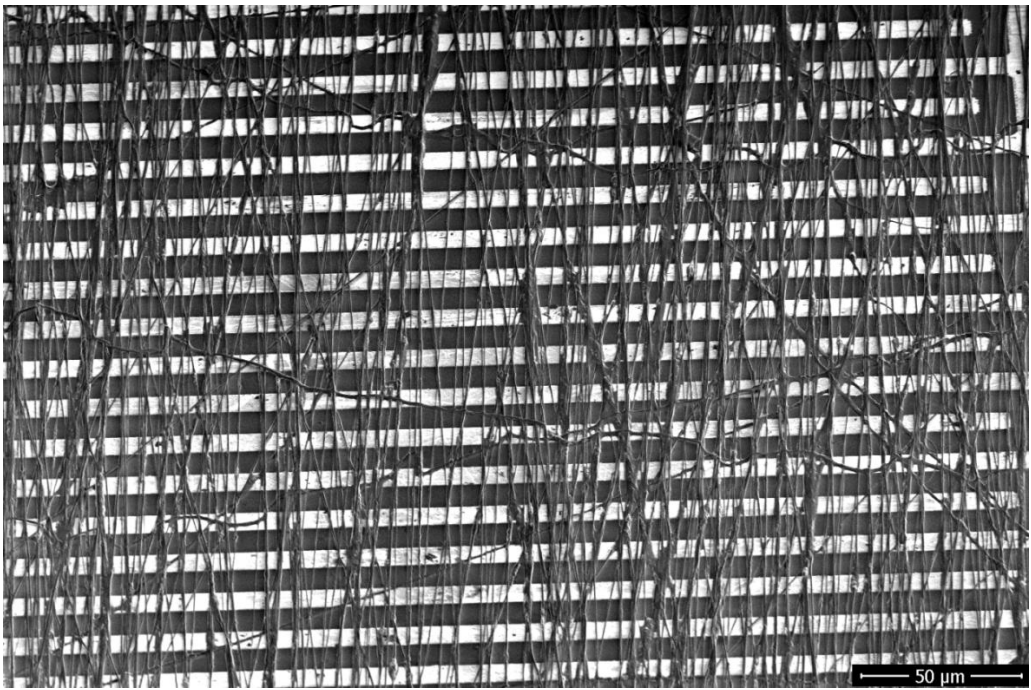

**Figure S2**. A monolayer of fibers electrospun onto the interdigitated electrodes.

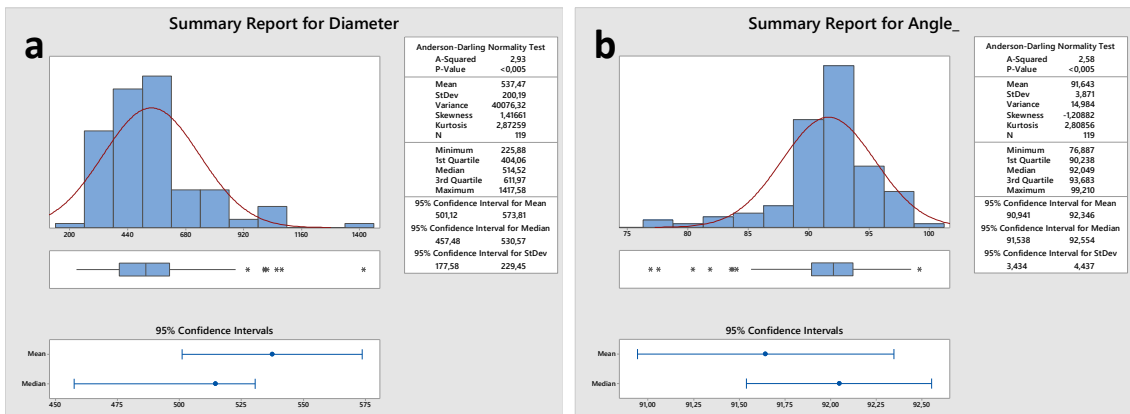

**Figure S3.** (a) Diameter distribution analysis based on 120 measurements and (b) Angle distribution analysis based on 120 measurements.

**Table S1.** Average diameter and angle of electrospun PCL-MWCNT/C60 fibers.

| **Fiber diameter (nm)** | 538 ± 18 |
|---|---|
| **Angle respect to the electrode fingers (°)** | 91.3 ± 0.4 |

**3. Raman spectra of MWCNT and C60/MWCNT isolated from fibers.** C60/MWCNT complexes were isolated from electrospun fibers by solvent extraction. The same extraction procedure was performed on fibers containing MWCNT only and the Raman spectra of both extracts were compared (Figure S4).

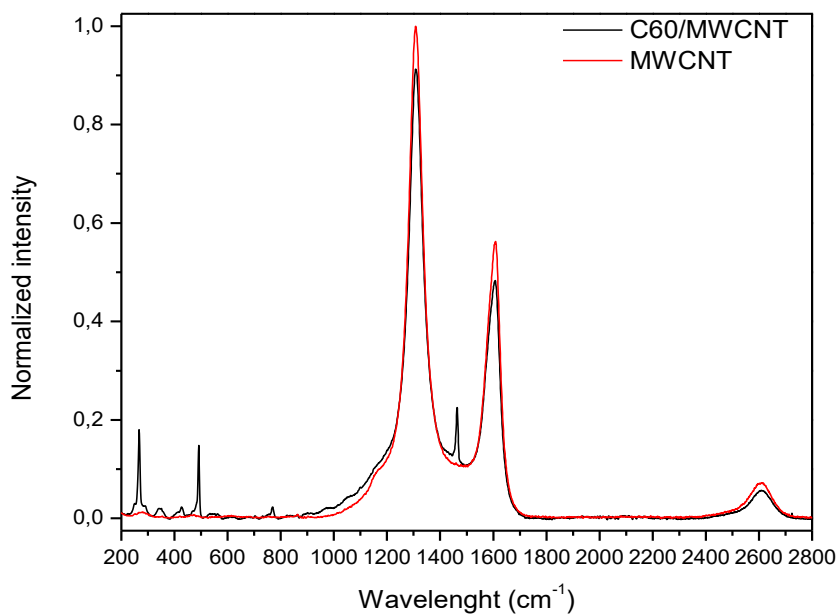

**Figure S4.** Raman spectra of C60/MWCNT complexes (black) and MWCNT (red) extracted from electrospun fibers.

**4. Thin layer chromatography.** In order to define if C60 could be effectively separated from the exohedral complex a thin layer chromatography was run on the extract. Mobile phase was toluene:hexane 8:2 (Figure S5).

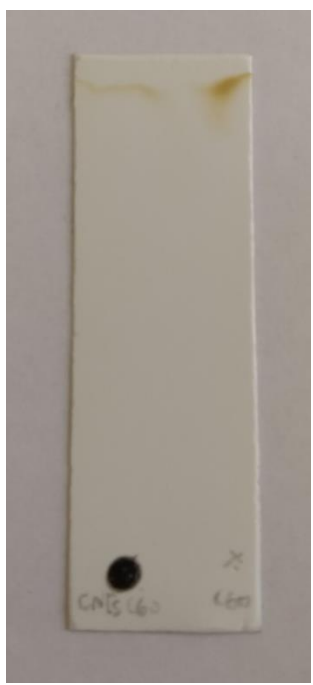

**Figure S5.** Thin layer chromatography of C60(right) and C60/MWCNT complexes extracted from the fibers(left).

**5. DSC plots.** Calorimetric studies were carried out in a TA Q2000 Differential Scanning Calorimeter (DSC) calibrated with indium. Ultra high purity nitrogen was used as a purge gas. Samples of approximately 9 mg each were encapsulated in aluminum pans and sealed. Cooling runs in order to determine the influence of MWCNTs, C60 and C60/MWCNT on crystallization temperature of PCL were performed (Figure S6).

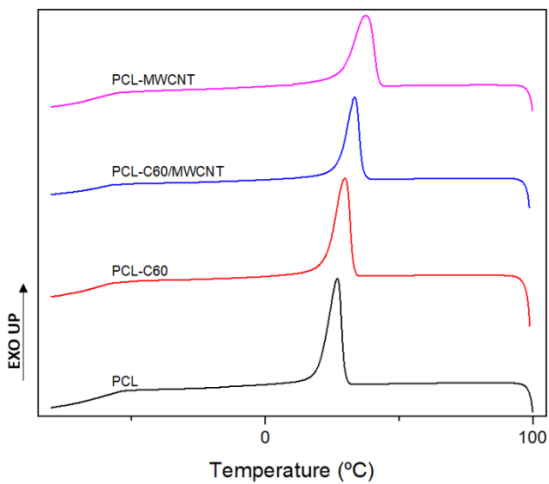

**Figure S6.** DSC curves showing crystallization peak of PCL, PCL-C60, PCL-C60/MWCNT and PCL-MWCNT fibers. An upshift for the crystallization temperature was observed upon the addition of C60, C60/MWCNT and MWCNT.

In order to calculate the effect of C60, MWCNT and C60/MWCNT on nucleation efficiency of PCL, a self-nucleation test was performed as follows:

The crystalline thermal history was erased by heating the samples at 100ºC for 3 min. Cooling and subsequent heating scans were registered at 10 ºC/min.

The thermal treatment comprises:

(a) Erasure of crystalline thermal history by heating the sample to 100 ºC for 3 min.

(b) Creation of a "standard" thermal history by cooling at 10 ºC/ min to 0 ºC.

(c) Partial melting up to a temperature denoted $T_s$.

(d) Thermal conditioning at $T_{s1}$ during 5 min.

(e) DSC cooling scan from $T_s$, in order to observe the effects of the thermal treatment will be reflected on the crystallization of the PCL.

(f) Complete melting heating at 10 ºC/min to 100 ºC aiming to observe a double melting peak.

(g) Cooling at10 ºC/min to 0 ºC.

(h) Thermal conditioning at $T_{s2}$ during 5 min.

(i) DSC cooling scan from $T_s$, in order to observe the effects of the thermal treatment will be reflected on the crystallization of the PCL.

(j) Complete melting heating at 10 ºC/min to 100 ºC aiming to observe a double melting peak.

This method was performed at different $T_s$ (70 ºC, 68 ºC, 65 ºC, 63 ºC, 61 ºC, 59 ºC, 57 ºC, 55 ºC) as shown in Figure S7.

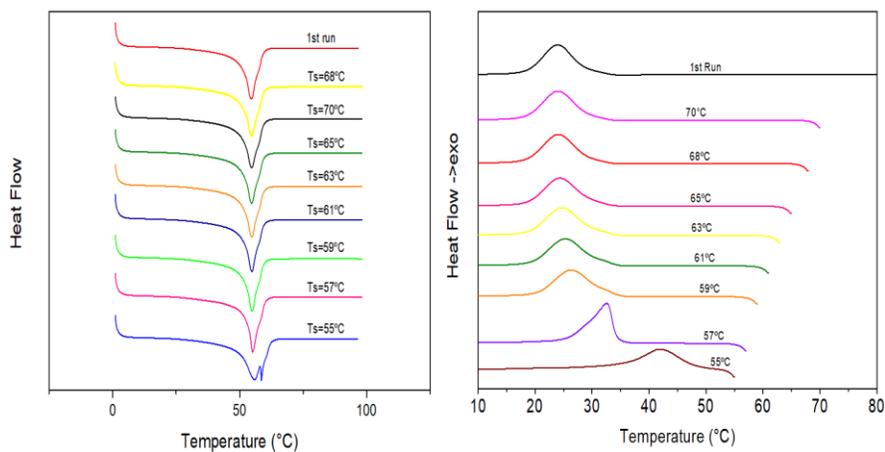

**Figure S7.** Heating runs (a) and cooling runs (b) of PCL after different partial melting temperatures ($T_s$).

**6. Fabrication and characterization of PCL-MWCNT/C60 spin-coated devices.** The devices were fabricated by spin coating the same dispersion used for electrospinning onto interdigitated gold electrodes of different width to length (W/L) ratio (L=1-100 µm, W/L=50-2000) patterned on a Si/SiO2 (300 nm) at 8500 rpm. Optical microscope images of the devices produced by electrospinning and spin coating is shown in Figure S8. The electrical stimulation and device reading plots are shown in Figures S9 a and b. The Raman spectrum of the spin-coated layer showing the pentagonal pinch mode of C60 at 1469 cm$^{-1}$ is shown in Figure S10.

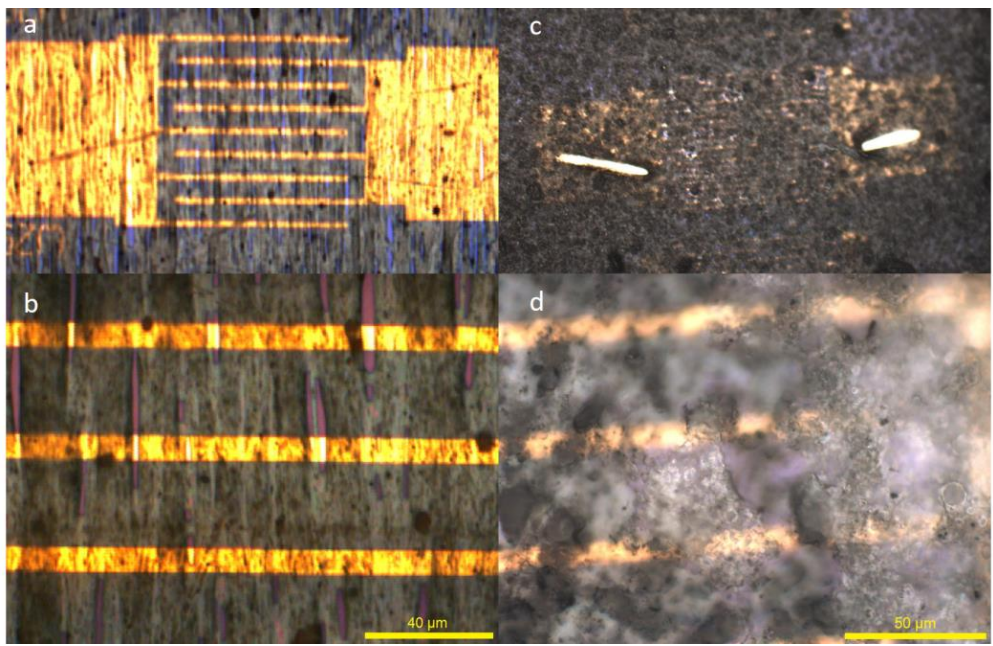

**Figure S8.** Optical microscope images of the devices fabricated by electrospinning (a and b) and spin coating (c and d)

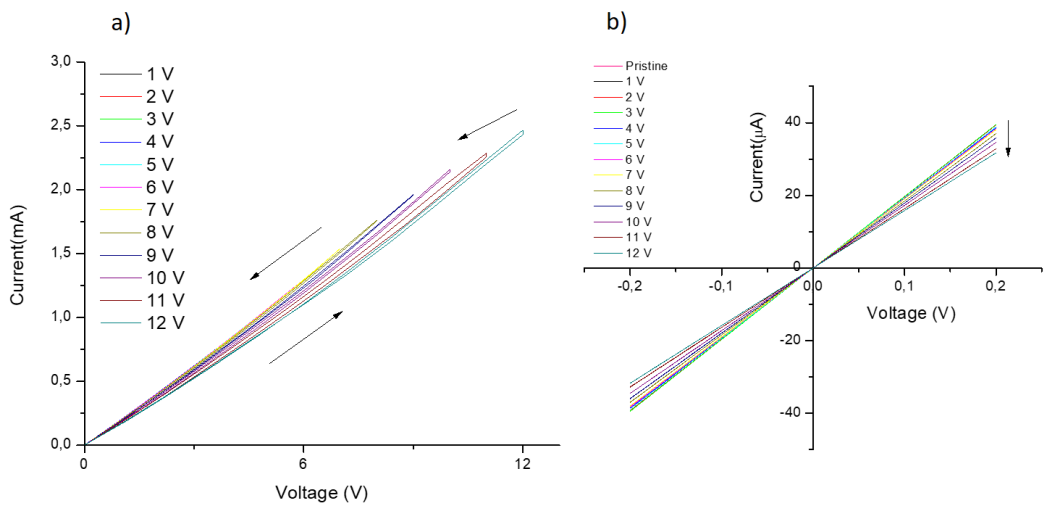

**Figure S9.** (a) Electrical stimulation of spin-coated memory devices starting from 0V with a step of 1V between voltage sweeps, (b) the electrical resistance estimated by the slope of the IV curve after each stimulus.

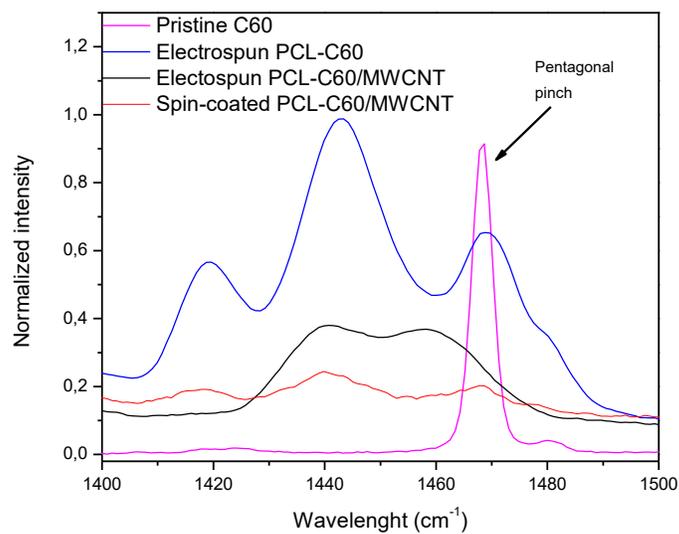

**Figure S10.** Raman spectra of electrospun fibers of PCL-C60(blue), electrospun fibers of PCL-C60/MWCNT(black), spin coated film of PCL-C60/MWCNT(red) and pristine C60(magenta).